\documentclass[final,authoryear,5p,times,twocolumn]{elsarticle}

\usepackage{amssymb}
\usepackage{epsfig}
 \usepackage{subfigure}
\usepackage{graphicx}
\def\astrobj#1{#1}

\journal{New Astronomy}

\begin{document}

\begin{frontmatter}



\title{Str\"omgren photometry and spectroscopy of the $\delta$ Scuti stars \astrobj{7 Aql} and \astrobj{8 Aql}}


\author[oan]{L. Fox Machado\corref{cor}}
\ead{lfox@astrosen.unam.mx}

\author[oan]{M. Alvarez}

\author[oan]{R. Michel}

\author[iaa]{A. Moya}

\author[ia]{J.H. Pe\~na}

\author[ia]{L. Parrao}

\author[oan]{A. Castro}

\cortext[cor]{Corresponding author. Tel.: +52 6461744580; fax +52
6461744607 }

\address[oan]{Observatorio Astron\'omico Nacional, Instituto de
Astronom\'{\i}a -- Universidad Nacional Aut\'onoma de M\'exico, Ap.
P. 877, Ensenada, BC 22860, M\'exico}

\address[iaa]{Instituto de Astrof\'{\i}sica de Andaluc\'{\i}a
(CSIC), PO Box 3004, 18080 Granada, Spain}

\address[ia]{Instituto de
Astronom\'{\i}a -- Universidad Nacional Aut\'onoma de M\'exico, Ap.
P. 70-264, M\'exico, D.F. 04510, M\'exico}

\begin{abstract}
Photometric $(ubvy-\beta$) and spectroscopic observations of the
$\delta$ Scuti variables \astrobj{7 Aql} and \astrobj{8 Aql} are
described. The Str\"omgren standard indices and physical parameters
of both stars are derived. Spectral types of F0V and F2III have been
assigned to \astrobj{7 Aql} and \astrobj{8 Aql} respectively
considering the results from both spectroscopy and photometry.
Differential $uvby$ light curves were also analyzed. A attempt of
multicolour mode identification is carried out.

\end{abstract}


\end{frontmatter}


\section{Introduction}
\label{sec:int}

\astrobj{7 Aql} (HD~174532, SAO~142696, HIP~92501) was discovered to
be a pulsating star of $\delta$ Scuti type by \cite{garrido1} in a
 search of new variables in preparation for the COROT
mission. Its multiperiodic nature was established by \cite{fox1} who
detected six oscillation frequencies. \astrobj{8 Aql} (HD~174589,
SAO~142706, HIP~92524) was reported as a new multiperiodic $\delta$
Scuti variable  with three pulsation frequencies by \cite{fox1}. The
dominant modes detected by \cite{fox1} were confirmed by \cite{fox3}
by using CCD photometry. Both stars represent interest for
asteroseismology since they are slightly evolved, and hence located
in the HR diagram in the ambiguous transition phase between core
hydrogen burning and thick shell hydrogen burning. This phase is
sensitive to the treatment of the core overshooting processes.
 Moreover, \astrobj{7 Aql} has been selected  as a secondary target of the
 COROT seismology program \citep{uytt}.
    The COROT space mission \citep{baglin}, successfully launched in
December 2006, is providing a huge number of detected oscillation
frequencies in individual $\delta$ Scuti stars
\citep{poretti1,garcia}. In order to fully exploit the asteroseismic
data by using stellar evolutionary models, accurate stellar physical
parameters are needed. For \astrobj{7 Aql} no $uvby-\beta$ indices
have been reported to date. On the contrary, a number of Str\"omgren
indices [$(b-y)$, $m_{1}$,
 $c_{1}$, $H_{\beta}$] have been reported
for  \astrobj{8 Aql}, but based on  a few measurements. Concerning
the spectral classification of the stars, the reported types are not
unique in the literature. Namely, the Michigan Catalogue of HD
stars, Vol.5 (Houk+, 1999)
 reports F0V and A9IV for \astrobj{8 Aql} and \astrobj{7 Aql}
 respectively. Whereas, the SAO Star Catalog J2000 (SAO
Staff 1966; USNO, ADC 1990) lists A2 for \astrobj{7 Aql} and A3 for
\astrobj{8 Aql}. The Bright Star Catalogue, 5th Revised Ed.
(Hoffleit+, 1991) gives F2III for \astrobj{8 Aql}. These
classifications are based mainly upon photographic spectra which are
less accurate than those obtained with modern equipments. The aim of
this paper is to present more precise information about \astrobj{7
Aql} and \astrobj{8 Aql} by using both Str\"omgren photometry and
spectroscopy. Furthermore, the differential time series in the
Str\"omgren bands $uvby$ are also analyzed.

\section{Observations and data reduction}

\subsection{Photometric observations}

The observations were secured in  2007 on the nights of June 21, 22,
23, 28, 30 and July 07 and 08 at the Observatorio Astr\'onomico
Nacional-San Pedro M\'artir (OAN-SPM), Baja California, Mexico. The
1.5-m  telescope with the six-channel Str\"omgren spectrophotometer
was implemented. The observing routine consisted of five 10 s of
integration of the star from which five 10 s of integration of the
sky was subtracted. Two constant comparison stars were observed as
well, namely HD 174046 and HD 174625. Along with \astrobj{7 Aql} and
\astrobj{8 Aql} also observed was the $\delta$ Scuti variables HD
170699. The results for this particular objects will be given
elsewhere \citep{alvarez1}.

A set of standard stars was also observed each night to transform
instrumental observations onto the standard system and to correct
for atmospheric extinction.  The instrumental magnitudes ($_{\rm
inst}$) and colours, once corrected from atmospheric extinction,
were transformed to the standard system ($_{\rm std}$) through the
well known transformation relations given by \cite{stromgren} :

\[ V_{\rm std} = A + y_{\rm inst} + B(b-y)_{\rm inst} \]

\[(b-y)_{\rm std} = C + D(b-y)_{\rm inst} \]

\[ m_{1,{\rm std}} = E + Fm_{1,{\rm inst}} + G(b-y)_{\rm inst} \]

\[ c_{1,{\rm std}} = H + Ic_{1,{\rm inst}} + J(b-y)_{\rm inst} \]

\[ H_{\beta,{\rm std}} = K + LH_{\beta,{\rm inst}} \]

\noindent where $V$ is the magnitude in the Johnson system, and the
$m_{1}$ and the $c_{1}$ indices are defined in the standard way:
$m_{1} \equiv (u-v) -(v-b)$ and $c_{1} \equiv (v-b) - (b-y)$.
 Applying the above
equations to the standard stars, an estimation of the mean errors
for the transformations to the standard system can be obtained:
   $\sigma_{v}= 0.011$,
$\sigma_{(b-y)} = 0.006$, $\sigma_{m_{1}}=0.015$,
$\sigma_{c_{1}}=0.015$, $\sigma_{H_{\beta}}=0.015$. The photometric
precision in the instrumental system was: $\sigma_{u}=0.017$,
$\sigma_{v}=0.013$, $\sigma_{b}=0.011$ $\sigma_{y}= 0.009$.

The averaged standard magnitudes and indices for target and
comparison stars  are given in Table~\ref{tab:index_pp}. The
Str\"omgren indices for \astrobj{7 Aql} and comparison stars are
reported for the first time in the present paper. Whereas, a number
of photometric indices [$(b-y)$, $m_{1}$,
 $c_{1}$, $H_{\beta}$] are available for \astrobj{8 Aql}. In
particular, \cite{crawford} gives (0.178, 0.178, 0.834, 2.747),
\cite{gronbech, gronbech1} list (0.176, 0.183, 0.831, 2.752) and
\cite{hauck} give (0.177, 0.181, 0.832, 2.749). These   are in
agreement with those reported in Table~\ref{tab:index_pp} within
1-$\sigma$ error.

\begin{table*}[!t]\centering
 \caption{Averaged standard magnitudes and indices for target
 and comparison stars.
The number of $uvby$ and $H_{\beta}$ measurements
 are indicated as
  $N_{uvby}$ and $N_{\beta}$ respective.
 }
  \label{tab:index_pp}
  \begin{tabular}{lcccccc}
\hline \hline
  Star&  $V$&  $(b-y)$&  $m_{1}$&  $c_{1}$& $H_{\beta}$& $N_{uvby}$/$N_{\beta}$ \\
& (mag)& (mag)& (mag)& (mag)& (mag)&\\
\hline
 \astrobj{7 Aql}  &  6.894 & 0.171   & 0.180  & 0.873 & 2.755 &  291/26  \\
 \astrobj{8 Aql}  &  6.075 & 0.178  & 0.185 & 0.822 & 2.730 & 288/26 \\
\astrobj{HD 174046} (c1) & 9.570 & 0.276 & 0.078 & 1.097 & 2.867 & 285/26\\
\astrobj{HD 174625} (c2) & 9.436 & 0.363 & 0.101 & 0.553 & 2.664 & 292/26\\
 \hline
\end{tabular}
\end{table*}

\subsection{Spectroscopic observations}\label{sec:spec_obs}
\begin{figure}[!t]
\centering
\includegraphics[width=7cm]{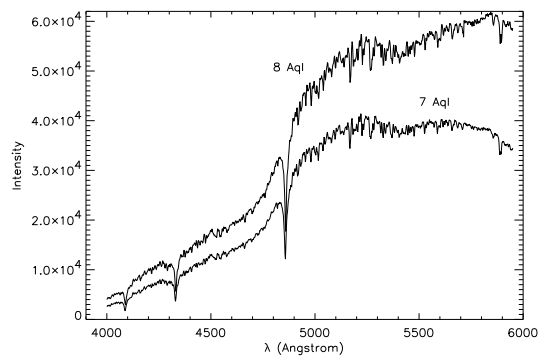}
\caption{Low resolution spectra of target stars \astrobj{8 Aql}  and
\astrobj{7 Aql}.} \label{fig:spectra}
\end{figure}

Spectroscopic observations were conducted at the 2.12-m telescope of
the same observatory during  July 24, 2008 (UT). We used the Boller
\& Chivens spectrograph installed in the Cassegrain focus of the
telescope. The 600 lines/mm gratting was used to cover a wavelength
range from 3900 to 6000 \AA. A
  dispersion of 2.05 \AA\, per
pixels with a resolution of 5.6 \AA\, was employed. The SITE3 $1024
\times 1024$ pixel CCD with a 0.24 $\mu$m pixel size was attached to
the spectrograph. Fig.~\ref{fig:spectra} displays examples of the
spectra, which were reduced in the standard way using the IRAF
package.
 A comparison of the
normalized spectra with those of well classified stars available in
the literature  was carried out. The spectrum of \astrobj{8 Aql} is
very similar to that of HD 89254 (F2III) of the library of stellar
spectra STELIB \citep{leborgne}. On the other hand, the spectrum of
the star HD 90277 (F0V) of the same library reproduces our spectra
of \astrobj{7 Aql} fairly well, besides that its resolution is twice
as high as our.

\section{Physical parameters}\label{sec:par}

 We have used the standard indices
$uvby-\beta$
 listed in Table~\ref{tab:index_pp}
to estimate the reddening as well as the unreddened colours of our
target stars. The calibrations of \cite{nissen} which are based on
calibrations of \cite{crawford, crawford2, crawford1} for A- and F-
type stars were implemented. The derived physical parameters  are
listed in Table~\ref{tab:nissen_par}.
 The confidence of the physical parameters can be assessed by comparing
the distances derived in the present study with those estimated from
accurate trigonometric parallaxes. In particular, the HIPPARCOS
parallax measurement for \astrobj{7 Aql} is $7.70 \pm 0.80$ mas
 and for \astrobj{8 Aql} is $11.80 \pm 0.78$ mas. The corresponding
 distances are $130 \pm 15$ pc and $85 \pm 6$ pc respectively. Thus,
 there is a good agreement between
trigonometric and photometric distances. The $T_{\rm eff}$, log $g$
and metallicity from observed colours have been determined by means
of the code TempLogG \citep{rogers, kupka}. The resulting physical
parameters  are listed in Table~\ref{tab:templog_par}.
 The spectral types and  luminosity classes of the stars can be determined
through the relationship  between MK spectral types and the
reddening free indices $\beta$, $[m_{1}]=m_{1}+0.18(b-y)$,
$[c_{1}]=c_{1}-0.20(b-y)$ by \cite{oblak}. Considering the indices
listed in Table~\ref{tab:index_pp} with the errors, we have found a
spectral classification for the target stars between A9 and F2, with
luminosity class of either III or V. Therefore, no unique spectral
type can be obtained from St\"omgren photometry. This is due to the
fact that the Str\"omgren standard indices used  to assign a
determined MK type in \citet{oblak} have a rather high standard
deviation, hence more than one MK type could be assigned to the
target stars.

\begin{table*}[!t]\centering
 \caption{Physical parameters  for  the target stars derived from Nissen's (1988) calibrations.}
\label{tab:nissen_par}
  \begin{tabular}{lcccccccr}
\hline \hline
 Star & $E(b-y)$& $(b-y)_{0}$&  $m_{0}$& $c_{0}$& $\beta_{0}$&$m_{V}$&$M_{V}$& Distance\\
&(mag)&(mag)&(mag)&(mag)&(mag) &(mag)&(mag)&(pc)\\
\hline
  \astrobj{7 Aql} & 0.000 &0.173 &  0.180 &  0.873  &2.755 &   6.89  & 1.25  &  134 \\
  \astrobj{8 Aql} &  0.000& 0.196 &  0.185 &  0.822 &2.730 &   6.07 &  1.27  &   92   \\

\hline
\end{tabular}
\end{table*}

\section{Differential light curves and frequency
analysis}\label{sec:difphot}

\begin{figure*}[!t]
\begin{center}
\includegraphics[width=14cm]{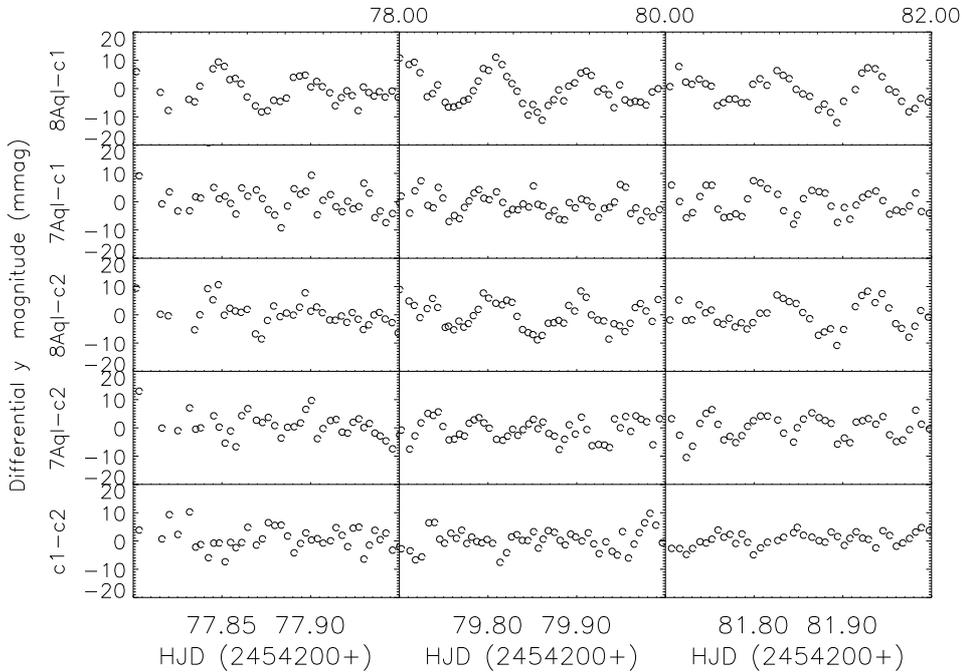}
\caption{Examples of the differential light curves taken with the
Str\"omgren spectrophotometer using the $y$ filter with reference
star HD 174046 $=$ c1 and HD 174625 $=$ c2. The name of each one
differential light curve is indicated at left.}
\label{fig:curves_pp}
\end{center}
\end{figure*}

 Figure~\ref{fig:curves_pp} displays
examples of the differential light curves in the $y$ Str\"omgren
filter of the target stars for three selected nights (vertical
panels). The last three horizontal panels, from left to right,
correspond to the differential light curve \astrobj{HD
174046}~-~\astrobj{HD 174625}.

A period analysis has been performed by means of standard Fourier
analysis and leas-squares fitting. In particular,  the amplitude
spectra of the differential time series were obtained by means of an
iterative sinus wave fit (ISWF; \citealt{ponman}).

\begin{figure*}[!t]
  \subfigure[]{\includegraphics[width=11cm]{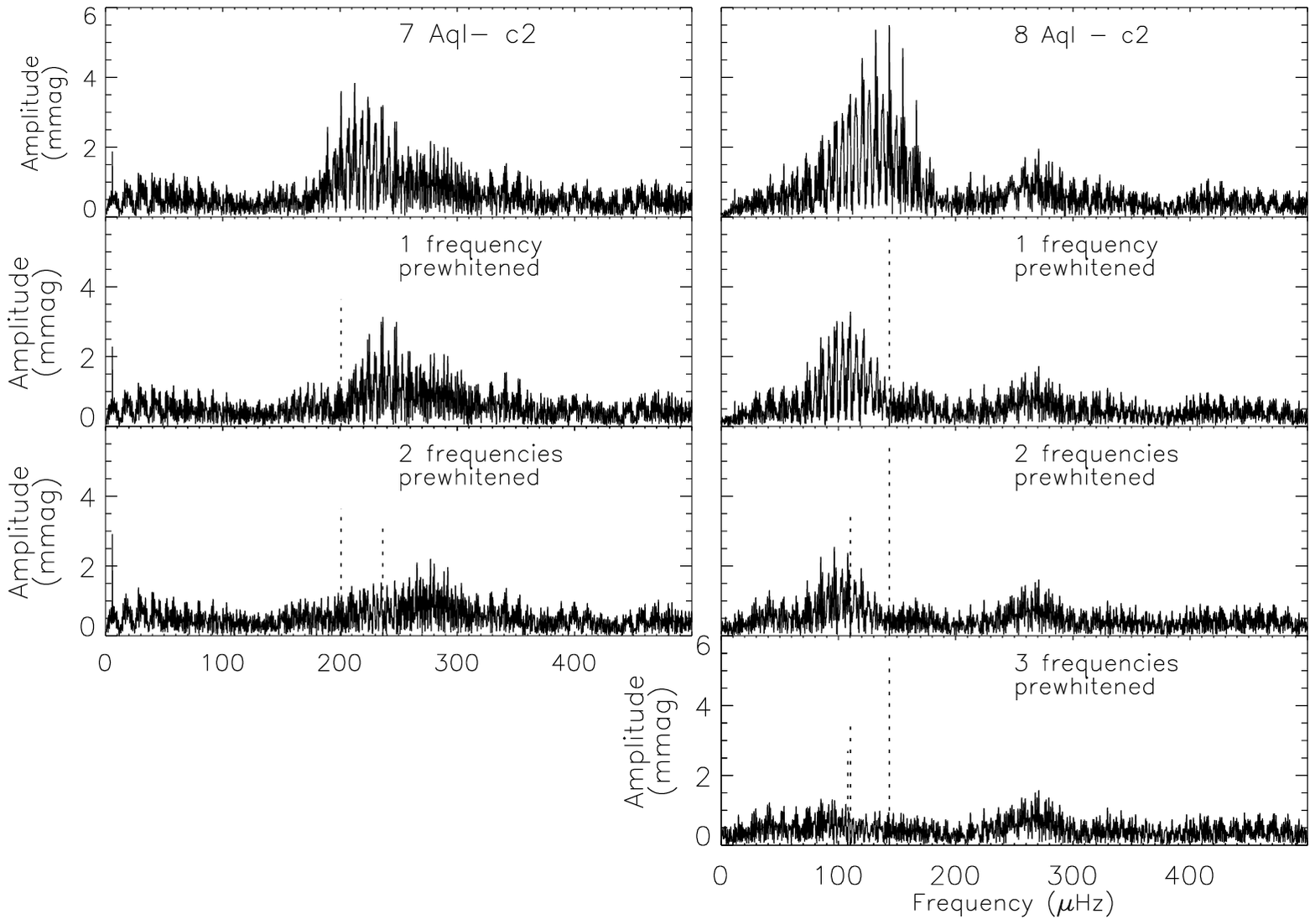}}
  \subfigure[]{\includegraphics[width=7cm]{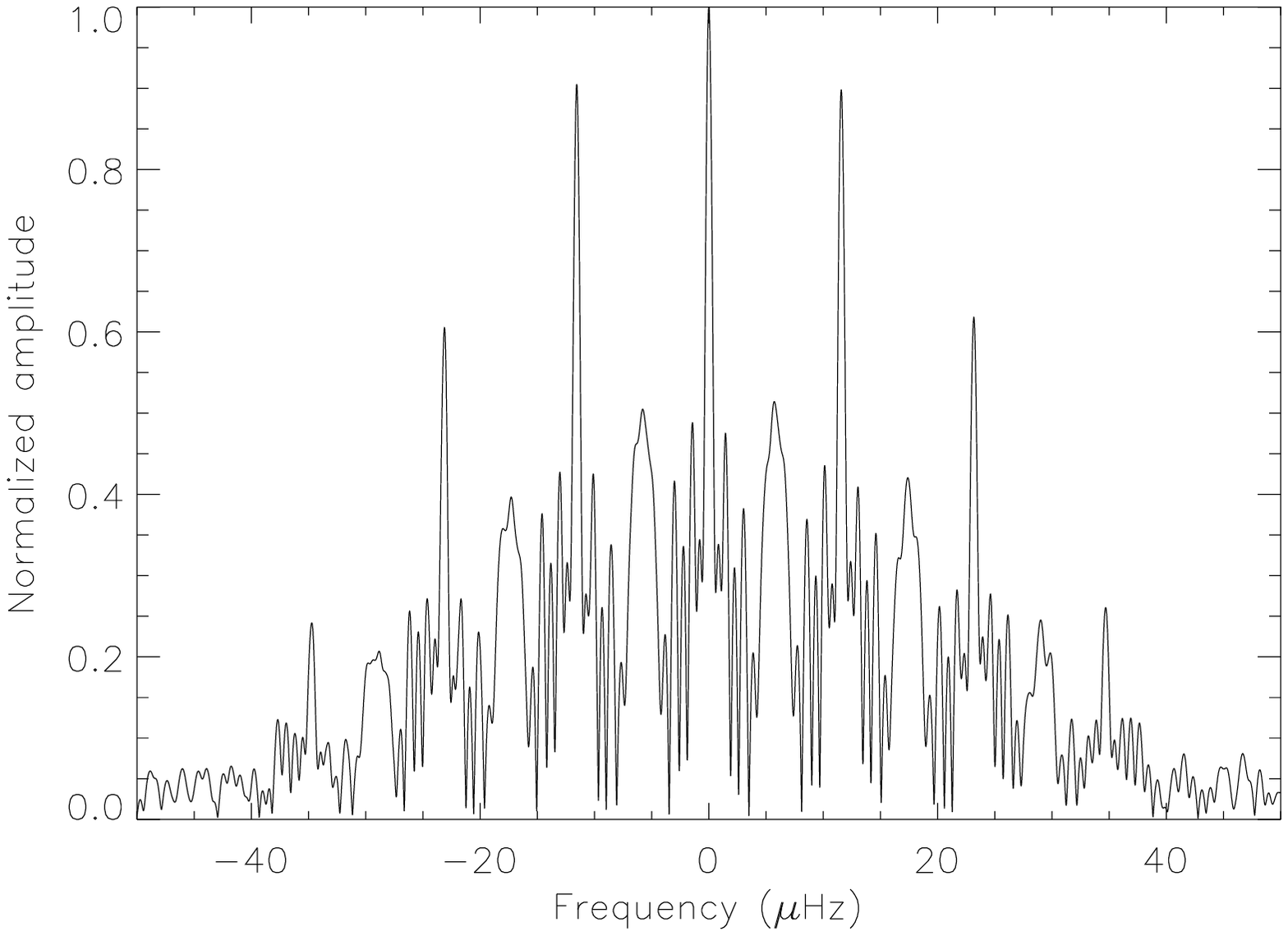}}
  \caption{(a) Pre-whitening process in the spectra derived from the
  PP $v$ differential light curves 7 Aql$-$c2 (left) and 8 Aql$-$c2 (right).
  (b) Spectral windows in amplitude}
  \label{fig:prewh_pp}
\end{figure*}

The amplitude spectra  of the differential  $v$ light curves
\astrobj{7 Aql}$-$c2, \astrobj{8 Aql}$-$c2  are shown in the top
panels of each plot of Figure~\ref{fig:prewh_pp}(a).  The subsequent
panels of each plot in the figure, from top to bottom, illustrate
the  process of detection of the frequency peaks in each amplitude
spectrum. We followed the same procedure as explained in
\citet{alvarez} and employed by \cite{fox4, fox, fox5, fox6}. We
have used a threshold of 3.7-$\sigma$ above  the mean noise level of
the 100 $\mu$Hz closest to the peak in the amplitude spectrum to
consider a frequency as statistically significant, as described in
\citet{alvarez}.

The windows function of the  observations is shown in
 ~\ref{fig:prewh_pp}(b).
  The resolution measured from the FWHM of the main
lobe in the spectral window is $\Delta\,\nu=1.1$ $\mu$Hz. The
results obtained from the prewhitening process of the Str\"omgren
$vby$ time series  are listed in Table~\ref{tab:frec1}, where the
detected frequencies with their corresponding amplitudes and phases
are given.  The data from the Str\"omgren $u$ band are omitted
hereafter for sake of clarity. The formal errors derived from
non-weighting fits are also listed. We note that for uncorrelated
observations like ours these uncertainties usually may underestimate
the real errors in amplitudes and phases.

 The detected frequencies in the  Str\"omgren $v$ band ($\lambda
= 4110\,$\AA, $\Delta \lambda = 170\,$\AA) can be compared
  with those reported by \cite{fox1} whose observations were
  obtained through a similar interferometric blue filter ($\lambda \approx 4200\,$\AA, $\Delta
\lambda \approx 190\,$\AA). Nonetheless, the time series analyzed by
\cite{fox1} are based on a multisite campaign and therefore the
final resolution is better. They  detected three frequency peaks in
\astrobj{8 Aql} namely 108.04 $\mu$Hz (4.1 mmag), 110.20 $\mu$Hz
(6.1 mmag), 143.36 $\mu$Hz (9.6 mmag). As can be seen in
Table~\ref{tab:frec1} we have detected the same frequencies in this
season, but with smaller oscillation amplitudes. We note, however,
that the amplitude ratio $A_{\nu_{1}} / A_{\nu_{2}}$ and
$A_{\nu_{1}} / A_{\nu_{3}}$ are almost the same in both studies. In
particular, the amplitude of the dominant mode, $\nu_{3}$, is 1.8
times smaller in our one-site observations. Even though, the
oscillation amplitudes are big enough to be detected in our run.

Regarding \astrobj{7 Aql} \cite{fox1}
 detected six significant frequency peaks namely 193.28 $\mu$Hz (2.8 mmag), 201.05 $\mu$Hz (3.8
mmag), 222.08 $\mu$Hz (3.6 mmag), 223.96 $\mu$Hz (3.4 mmag), 236.44
$\mu$Hz (6.1 mmag), 295.78 $\mu$Hz (1.5 mmag).
  Among these we have
confirmed only two frequency peaks.  $\nu_{1}$ most likely
corresponds to 201.05 $\mu$Hz of \cite{fox1} with similar amplitude,
while $\nu_{2}$ beyond no doubt is the dominant mode 236.44 $\mu$Hz
with smaller oscillation amplitude. From the amplitude ratio of the
detected modes in \cite{fox1} a smaller amplitude is expected for
$\nu_{1} \sim 201$ $\mu$Hz. Therefore, the amplitude of this peak
probably was affected by the side lobes. We think  that the
difference in amplitude of the modes in both seasons is a
consequence of the bad coverage rather than intrinsic amplitude
variability. In fact, if we define the oscillation amplitude
($A_{\rm osc}$) as the maximum constructive interference of the
observed modes a short time series might induce an underestimation
of the amplitude especially in presence of beating phenomena due to
close frequencies. As shown by \cite{fox1} the oscillation
amplitudes of the modes in \astrobj{8 Aql} are on average 1.9 times
larger that those of \astrobj{7 Aql}. This explains the fact that we
have only detected two oscillation modes in \astrobj{7 Aql}, but all
in \astrobj{8 Aql}.

\begin{table*}[]\centering
 \caption{Fundamental parameters for the target stars computed
 with the TempLogG code.}
\label{tab:templog_par}
  \begin{tabular}{lccccr}
\hline \hline
 Star & $M_{V}$& D &$T_{\rm eff}$ &$\log g$& [Fe/H]\\
&(mag)&(pc)&(K)&&\\
\hline
  \astrobj{7 Aql} & 1.22  &  136 & 7257 &3.62 &0.01  \\
  \astrobj{8 Aql} & 1.23 &   92 & 7051  &  3.51  &0.14    \\
\hline
\end{tabular}
\end{table*}

\begin{table}[!t]\centering
  \setlength{\tabcolsep}{1.0\tabcolsep}
 \caption{Frequency peaks detected in the light curves \astrobj{7 Aql}~$-$~c2 and \astrobj{8 Aql}~$-$~c2. S/N
is the signal-to-noise ratio in amplitude after the prewhitening
process.  The origin of $\varphi$ is at 24544272.72047}
\label{tab:frec1}
  \begin{tabular}{ccccr}
\hline
$\nu$&Freq.&  A & $\varphi$ & $S/N$  \\
&($\mu$Hz)&(mmag)&(rad)&\\
\hline
&&\astrobj{7 Aql}  &&\\
&Filter $v$&&&\\
$\nu_{1}$ &$200.90 \pm 0.05$ &   $3.64 \pm 0.3$ &   $ -0.53 \pm 0.03$ &  5.9\\
$\nu_{2}$ &$236.53 \pm 0.05$ &   $3.22 \pm 0.3$ &  $  +0.10 \pm 0.03$  &  4.2\\
&Filter $b$&&&\\
 $\nu_{1}$&$200.91 \pm 0.04$ &  $3.38 \pm 0.3$ & $ -0.57 \pm 0.03$  & 6.1\\
 $\nu_{2}$&$236.55 \pm 0.06$ &  $2.48 \pm 0.3$ & $  +0.06  \pm 0.04$   & 3.8\\
&Filter $y$&&&\\
 $\nu_{1}$&$200.89 \pm 0.05$ &   $2.64 \pm 0.3$ & $ -0.50 \pm 0.03$  &  5.6\\
 $\nu_{2}$&$236.54 \pm 0.06$ &   $2.22 \pm 0.3$ & $ +0.23 \pm 0.04$  & 4.2\\
\hline
&&\astrobj{8 Aql} &&\\
&Filter $v$&&&\\
 $\nu_{1}$&$143.38 \pm 0.04$ &  $5.38 \pm 0.3$ & $-2.54 \pm 0.02 $ &   11.5 \\
 $\nu_{2}$&$110.24  \pm 0.05$ &  $3.56 \pm 0.3$ & $ -0.67 \pm 0.03$&  6.7 \\
  $\nu_{3}$&$107.99 \pm 0.06$ &   $2.69 \pm 0.3$ & $ +2.48 \pm 0.04$ & 5.1\\
&Filter $b$&&&\\
$\nu_{1}$&$143.37 \pm 0.04$ &  $4.92 \pm 0.3$ & $-2.42 \pm 0.02$ &   13.3\\
 $\nu_{2}$&$110.26 \pm 0.04$ &  $2.86  \pm 0.3$ & $-0.74 \pm 0.03$&  7.0\\
  $\nu_{3}$&$107.96 \pm 0.06$ &  $2.33  \pm 0.3$ & $+2.59 \pm 0.04$&  5.7   \\
&Filter $y$&&&\\
 $\nu_{1}$&$143.36 \pm 0.03$ & $ 3.98 \pm 0.3$ & $ -2.32 \pm 0.03$&   10.8\\
 $\nu_{2}$&$110.23 \pm 0.04$ & $ 1.80 \pm 0.3$ & $ -0.44 \pm 0.05$&   4.3\\
  $\nu_{3}$&$107.92 \pm 0.05$ & $ 1.49 \pm 0.3$ & $ +2.91 \pm 0.06$&  3.6\\
\hline
\end{tabular}
\end{table}

\section{Preliminary comparison with theoretical models}\label{sec:models}

In this section the pulsation constants will be computed in order to
try to disentangle possible radial modes. Then the frequencies
listed in Table~\ref{tab:frec1} will be used in an attempt of
multicolour mode identification. A  more complete modelling
considering the frequency modes detected by \cite{fox1} will be
given in a forthcoming paper.

\begin{figure}[!t]
\centering
\includegraphics[width=7cm]{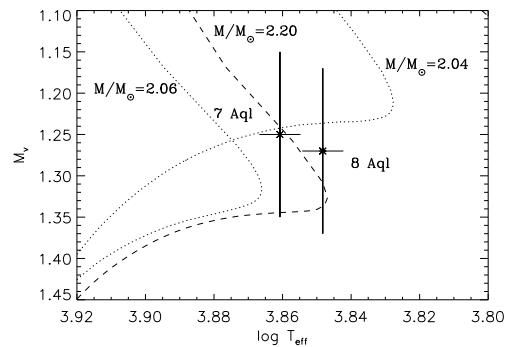}
\caption{HR diagram showing the location of the target stars. The
slightly cooler star is \astrobj{8 Aql}. Evolutive sequences of
non-rotating models without overshooting are shown by dotted ([Fe/H]
= 0.066) and dashed lines ([Fe/H] = 0.148).  The error bars give the
position of the stars according to the uncertainties discussed in
Sect~\ref{sec:par}.} \label{fig:models}
\end{figure}

 Figure~\ref{fig:models} shows the de-reddened position of the
target stars in an $T_{\rm eff}$-magnitude diagram. The computation
of the theoretical evolutive sequences are explained in
\citet{fox2}. The observed absolute magnitudes $M_{V}$  were taken
from Table~\ref{tab:nissen_par}, while the $T_{\rm eff}$ are from
Table~\ref{tab:templog_par}. Error bars of 0.1 mag for $M_{V}$ and
100 K for $T_{\rm eff}$ have been adopted. The dotted lines are
evolutive sequences of non-rotating models without overshooting
 giving a range of masses suitable for \astrobj{7 Aql}. The
dashed line corresponds to an evolutive track of models of 2.20
$M_{\odot}$  which match approximately the observational position of
\astrobj{8 Aql}. We have used a chemical initial composition of
[Fe/H] = 0.066 for \astrobj{7 Aql} and [Fe/H] = 0.148 for \astrobj{8
Aql}. According to the models depicted in Fig.~\ref{fig:models} the
mass of \astrobj{8 Aql} is 0.2 $M_{\odot}$ larger  than that of
\astrobj{7 Aql}. Their ages should be between 800 and 1000 Myr.  We
note that the effect of rotation have been neglected.  However as
shown by \cite{perez} the effect of rotation is important not only
in the location of the stars in a colour-magnitude diagram  but also
on the pulsation modes even for low rotators like \astrobj{7 Aql}
($v\,sin\,i=32$ km/s).

The pulsation constant $Q$ is expressed in terms of four observable
parameters as follows \citep{breger1}:

\begin{equation}
\log Q = -6.456 + \log P + 0.5 \log g + 0.1 M_{\rm bol} + \log
T_{\rm eff}
\end{equation}

 Using the parameters listed
in Table~\ref{tab:templog_par} and considering the balometric
corrections for the target stars \citep{balona}, we find for
\astrobj{7 Aql} $Q_{\nu_{1}}=0.0127$ and  $Q_{\nu_{2}}=0.0108$. For
\astrobj{8 Aql} we have $Q_{\nu_{1}}=0.0153$,  $Q_{\nu_{2}}=0.0199$
and $Q_{\nu_{3}}=0.0203$.
 Comparing
these $Q$-values with the theoretical ones \citep[2.0M48 model by
][]{fitch} we find that the oscillation modes $\nu_{1}$ and
$\nu_{2}$ of \astrobj{7 Aql} are indicative of $p$ modes of either
$l=0, 2\; {\rm or}\; 3$ with overtones $n=5$ and $n=7$,
respectively.

On the other hand, the 2.0M49 model \citep{fitch}, which
 match approximately the parameters of \astrobj{8 Aql} yields either identifications
$(l=1, n=4)$ or $(l=2, n=4)$  for  $\nu_{1}$, while the frequencies
$\nu_{2}$ and $\nu_{3}$ are consistent  with either radial
pulsations $(l=0, n=2)$ or non-radial oscillations $(l=2, n=2)$.

\begin{figure*}[!t]
  \subfigure[]{\includegraphics[width=7cm]{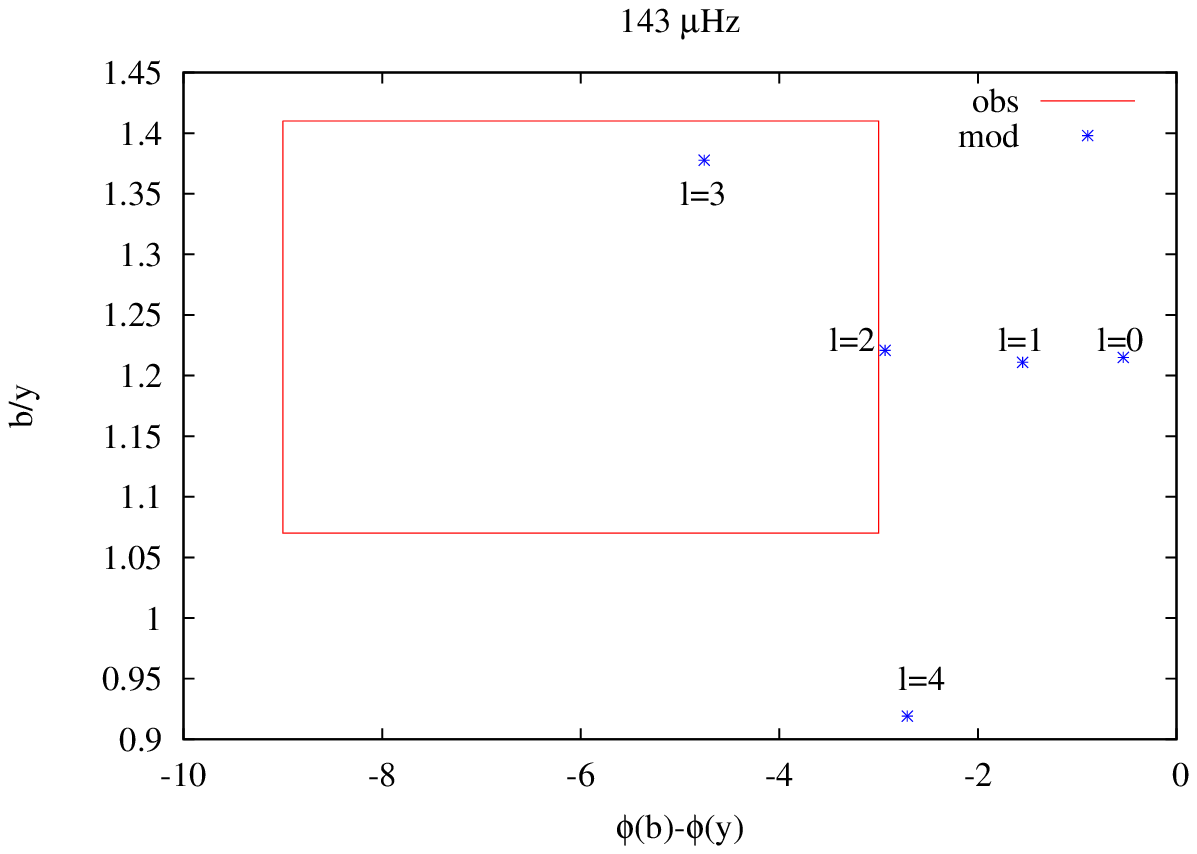}}
  \subfigure[]{\includegraphics[width=7cm]{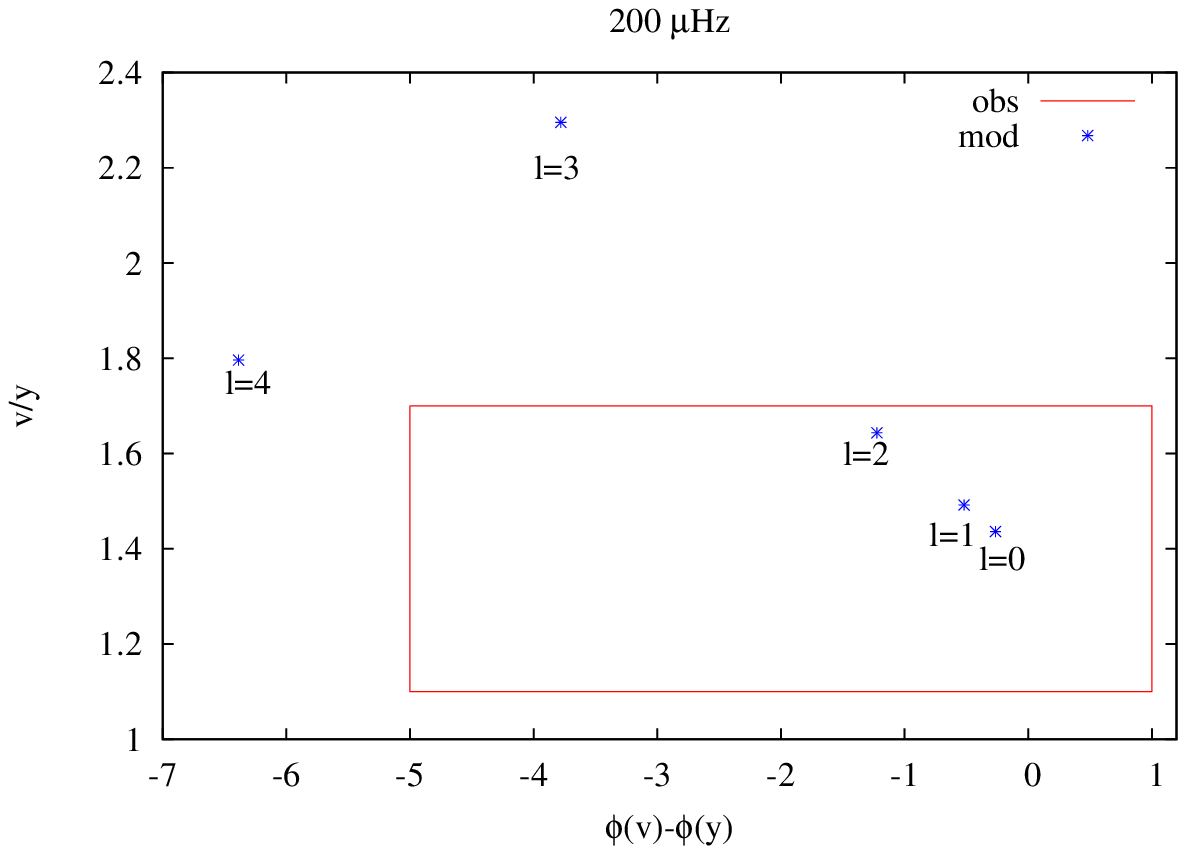}}
  \caption{Phase-amplitude diagram for the Str\"omgren filters showing
  the comparison between observed difference phase-amplitude ratio (boxes) and theoretical predictions
  (asterisks).
(a) $b$ and $y$ bands for the frequency $\nu_{1}=143.38\,\mu$Hz of
\astrobj{8 Aql}. (b) $v$ and $y$ bands for the frequency
$\nu_{1}=200.90\,\mu$Hz of \astrobj{7 Aql}. }
  \label{fig:phase}
\end{figure*}

\subsection{Multicolour photometry}

It is well known that the amplitude and phases observed with the
Str\"omgren filters can be used to estimate the spherical degree $l$
of the mode \citep{watson,garrido}. This estimation is done
comparing the observed amplitude ratios and phase differences with
theoretical predictions.

One equilibrium model per star has been obtained, passing by the
center of the observed photometric error box. The numerical code
CESAM \citep{morel} is used to obtain these models, fixing standard
physics for $\delta$ Scuti stars \citep[see][]{casas}. The
non-adiabatic pulsational code GraCo \citep{moya04,moya08} has also
been used for obtaining the variation of the flux and the phase-lags
necessary to compute the variation of the magnitude as a function of
the wave length. As both stars are close to the red edge of the
$\delta$ Scuti's instability strip, the outer convection zone is
well developed. Therefore, the equations describing the
convection-pulsation interaction are required. To do so, the Time
Dependent Convection \citep{grigahcene} has been used in GraCo.

Examples of the final comparison between observations and
theoretical predictions are depicted in Fig.~\ref{fig:phase}. The
horizontal  axe shows the phase difference in gradus, while the
vertical one the amplitude ratios.

Unfortunately, with the large error bars derived for the present
sparse observations, in some cases the oscillations are compatible
with all possible non-radial and radial modes up to $l=3$. In
others, the discrimination is not good. Even so, most
identifications point towards the presence of degrees with $l \geq
2$ values. If that hypothesis were correct the presence of radial
oscillations derived from Fitch's models could be excluded.
Therefore, the two detected frequencies in \astrobj{7 Aql} could be
identified as $l=2$ and $n=5, 7$; while the three frequencies in
\astrobj{8 Aql} would be consistent with $l=2$ and $n=4, 2, 3$.
However, continuous multicolour time series are required for a more
conclusive study.

\section{Conclusions}

We have presented the results obtained in an one-site observational
photometric campaign on  the $\delta$ Scuti stars \astrobj{7 Aql}
and \astrobj{8 Aql}. Photoelectric photometric $uvby-\beta$ data
were acquired at the 1.5-m telescope of SPM observatory by using the
Str\"omgren six channel spectrophotometer.

Str\"omgren standard indices for \astrobj{7 Aql}, \astrobj{8 Aql}
and comparison stars are reported.
 The main physical parameters have been derived from
the Str\"omgren photometry. These have been used to place the target
stars in an $T_{\rm eff}$-magnitude diagram. The star 8 Aql is about
0.2 $M_{\odot}$ more massive than 7 Aql, while their absolute
magnitudes are rather similar.

The pulsation constant $Q$ has been computed for the modes detected
in the present study.  An attempt of mode identification by means
multicolour photometry has been carried out. The stars seem to
oscillate with modes of degree $l=2$ or higher. However,  longer
multicolour time series are required for a more conclusive study.
 The analysis of few low
resolution spectra points that 7 Aql and 8 Aql are  stars of
spectral type F0V and F2III respectively.

\bigskip
{\bf \noindent Acknowledgments}

This work has received financial support from the UNAM under grant
PAPIIT  IN108106 and IN114309. A. M. acknowledges financial support
from a ``Juan de la Cierva'' contract of the Spanish Ministry of
Education and Science. Special thanks are given to the technical
staff and night assistant of the San Pedro M\'artir Observatory.
This research has made use of the SIMBAD database operated at CDS,
Strasbourg (France).

\bigskip
{\noindent \bf References}
\medskip

\bibliographystyle{elsarticle-harv}

\begin{thebibliography}{00}

\bibitem[\protect\citeauthoryear{\'Alvarez et al.}{1998}]{alvarez} \'Alvarez, M., et al. 1998, A\&A, 340, 149

\bibitem[\protect\citeauthoryear{\'Alvarez et al.}{2009}]{alvarez1} \'Alvarez, M., et al. 2009, RevMexAA (SC), 35,
148

\bibitem[\protect\citeauthoryear{Baglin et al.}{2006}]{baglin} Baglin, A., et al. 2006, in:
Proceedings of SOHO 18/GONG 2006/HELAS I, Beyond the spherical Sun,
ESA SP, 1306, 33

\bibitem[\protect\citeauthoryear{Balona}{1994}]{balona} Balona, L.A., 1994, MNRAS, 268,
119

\bibitem[\protect\citeauthoryear{Breger}{1990}]{breger1} Breger, M. 1990, Delta Scuti
Newsletter, 2, 13


\bibitem[\protect\citeauthoryear{Casas et al.}{2009}]{casas} Casas, R., et al. 2009, ApJ, 697,
522

\bibitem[\protect\citeauthoryear{Crawford}{1966}]{crawford2} Crawford, D.L. 1966, AJ, 71,
709

\bibitem[\protect\citeauthoryear{Crawford}{1975}]{crawford} Crawford, D.L. 1975, AJ, 80, 955

\bibitem[\protect\citeauthoryear{Crawford}{1979}]{crawford1} Crawford, D.L. 1979, AJ, 84,
185

\bibitem[\protect\citeauthoryear{Fitch}{1981}]{fitch} Fitch, W.S. 1981, ApJ, 249,
218

\bibitem[\protect\citeauthoryear{Fox Machado et al.}{2000}]{fox4} Fox Machado, L., et al. 2000,
in The Impact of Large Scale Surveys on Pulsating Stars Research,
eds. L. Szabados \& D. Kurtz, ASP Conf. Ser. 203, 477



\bibitem[\protect\citeauthoryear{Fox Machado et al.}{2002a}]{fox} Fox Machado, L., et al. 2002a, A\&A, 382, 556

\bibitem[\protect\citeauthoryear{Fox Machado et al.}{2002b}]{fox5} Fox Machado, L., et al. 2002b, in Radial and Nonradial
Pulsations as Probes of Stellar Physics,
 eds. C. Aerts, T. Bedding \& J. Christensen-Dalsgaard, ASP Conf. Ser. 259, 338

\bibitem[\protect\citeauthoryear{Fox Machado et al.}{2006}]{fox2} Fox Machado, L., et al. 2006, A\&A, 446, 611


\bibitem[\protect\citeauthoryear{Fox Machado et al.}{2007}]{fox1} Fox Machado, L., et al. 2007, AJ,
134, 860




\bibitem[\protect\citeauthoryear{Fox Machado et al.}{2008a}]{fox3} Fox Machado, L., et al. 2008a,
CoAst, 153, 20



\bibitem[\protect\citeauthoryear{Fox Machado et al.}{2008b}]{fox6} Fox Machado, L., et al. 2008b,
CoAst, 157, 307


\bibitem[\protect\citeauthoryear{Garc\'{\i}a Hern\'andez et al.}{2009}]{garcia} Garc\'{\i}a Hern\'andez, A., et al.
2009, A\&A, 506, 79

\bibitem[\protect\citeauthoryear{Garrido et al.}{1990}]{garrido} Garrido, R., et al. 1990, A\&A, 234,
262

\bibitem[\protect\citeauthoryear{Garrido et al.}{2002}]{garrido1} Garrido, R., et al. 2002,
in: {\it New $\delta$ Sct and $\gamma$ Dor variables in the COROT
field-of-view (Center direction)} - Version Li\'egeoise, CW3,
Li\'ege, December 2002


\bibitem[\protect\citeauthoryear{Grigahc{\`e}ne et al.}{2005}]{grigahcene}Grigahc{\`e}ne,
A., et al. 2005, A\&A, 434, 1055


\bibitem[\protect\citeauthoryear{Gronbech \& Olsen}{1976}]{gronbech} Gronbech, B., Olsen, E.H., 1976, AAS,
25, 213

\bibitem[\protect\citeauthoryear{Gronbech \& Olsen}{1977}]{gronbech1} Gronbech, B., Olsen, E.H., 1977, AAS,
27, 443

\bibitem[\protect\citeauthoryear{Hauck \& Mermilliod}{1976}]{hauck} Hauck, B., Mermilliod, M., AAS, 129,
431


\bibitem[\protect\citeauthoryear{Kupka \& Bruntt}{2001}]{kupka} Kupka, F., Bruntt, H.
2001, in Sterken C. ed., First COROT/MONS/MOST Ground Support
Workshop. Vrije Universiteit Brussel, Brussel, p. 3



\bibitem[\protect\citeauthoryear{Le Borgne et al.}{2003}]{leborgne} Le Borgne, J.-F., et al. 2003,
A\&A, 402, 433


\bibitem[\protect\citeauthoryear{Morel}{1997}]{morel} Morel, P. 1997, A\&AS, 124, 597


\bibitem[\protect\citeauthoryear{Moya et al.}{2004}]{moya04}Moya, A., Garrido, R., \& Dupret,
M.A.\ 2004, A\&A, 414, 1081

\bibitem[\protect\citeauthoryear{Moya \& Garrido}{2008}]{moya08} Moya, A., Garrido, R.
2008, Ap\&SS, 316, 129

\bibitem[\protect\citeauthoryear{Nissen}{1988}]{nissen} Nissen, P. 1988, A\&A, 199,
146

\bibitem[\protect\citeauthoryear{Oblak et al.}{1975}]{oblak} Oblak, E., et al. 1975, A\&AS, 24,
69

\bibitem[\protect\citeauthoryear{P\'erez Hern\'andez, et al.}{1999}]{perez} P\'erez Hern\'andez, F., et al. 1999, A\&A,
346, 586

\bibitem[\protect\citeauthoryear{Ponman}{1981}]{ponman} Ponman T. 1981, MNRAS, 196, 543


\bibitem[\protect\citeauthoryear{Poretti et al.}{2009}]{poretti1} Poretti, E., et al. 2009,
A\&A, 506, 85

\bibitem[\protect\citeauthoryear{Rogers}{1995}]{rogers} Rogers, N.Y. 1995, CoAst, 78,
1

\bibitem[\protect\citeauthoryear{Str\"omgren}{1966}]{stromgren}Str\"omgren, B. 1996, ARA\&A, 4, 433

\bibitem[\protect\citeauthoryear{Uytterhoeven et al.}{2009}]{uytt}Uytterhoeven, K., et al., 2009,
AIP Conf. Proc., 1170, 327



\bibitem[\protect\citeauthoryear{Watson}{1988}]{watson} Watson, R.D. 1988, Ap\&SS,
140, 255


\end{thebibliography}

\end{document}